\documentclass[preprint, sort&compress]{elsarticle}
\journal{Computer Physics Communications}

\usepackage[plain]{algorithm}
\usepackage[rightComments=false, endLComment=$$, commentColor=mydarkgray]{algpseudocodex}
\usepackage{amsmath}
\usepackage{amssymb}
\usepackage{bm}
\usepackage[nolabel]{showlabels}
\usepackage[ISO=false]{diffcoeff}
\usepackage{listings}
\usepackage{graphicx}
\usepackage[utf8]{inputenc}
\usepackage{tablefootnote}
\usepackage{threeparttable}
\usepackage{makecell}
\usepackage{xcolor}
\definecolor{mydarkgray}{gray}{0.2}

\usepackage{geometry}

\renewcommand{\b}[1]{\mathbf{#1}}

\begin{document}

\begin{frontmatter}

\title{Hierarchical autoregressive neural networks for statistical systems}
\author[iis]{Piotr Białas}
\ead{piotr.bialas@uj.edu.pl}
\author[ift]{Piotr Korcyl}
\ead{piotr.korcyl@uj.edu.pl}
\author[ift]{Tomasz Stebel}
\ead{tomasz.stebel@uj.edu.pl}

\address[iis]{Institute of Applied Computer Science, Jagiellonian University, ul.~Łojasiewicza 11, 30-348 Kraków Poland}
\address[ift]{Institute of Theoretical Physics, Jagiellonian University, ul.~Łojasiewicza 11, 30-348 Kraków Poland}

\begin{abstract}
It was recently proposed that neural networks could be used to approximate many-dimensional probability distributions that appear e.g. in lattice field theories or statistical mechanics. Subsequently they can be used as variational approximators to asses extensive properties of statistical systems, like free energy, and also as neural samplers used in Monte Carlo simulations. 
The practical application of this approach is unfortunately  limited by its unfavorable scaling both of the numerical cost required for training, and the memory requirements with the system size. This is due to the fact that the original proposition involved a neural network of width which scaled with the total number of degrees of freedom, e.g.  $L^2$ in case of a two dimensional $L\times L$ lattice. In this work we propose a hierarchical association of physical degrees of freedom, for instance spins, to neurons which  replaces it with the scaling with the linear extent $L$ of the system. We demonstrate our approach on the two-dimensional Ising model by simulating lattices of various sizes up to $128 \times 128$ spins, with time benchmarks reaching lattices of size $512 \times 512$. We observe that our proposal improves the quality of neural network training, i.e. the approximated probability distribution is closer to the target that could be previously achieved. As a consequence, the variational free energy reaches a value closer to its theoretical expectation and, if applied in a Markov Chain Monte Carlo algorithm, the resulting autocorrelation time is smaller. Finally, the replacement of a single neural network by a hierarchy of smaller networks considerably reduces the memory requirements.
\end{abstract}


\end{frontmatter}

\section{Introduction}

In Markov Chain Monte Carlo simulations the main obstacle in improving the statistical precision are the autocorrelations between consecutive samples. The recent idea \cite{PhysRevD.100.034515, 2020PhRvE.101b3304N} to alleviate this problem is based on the introduction of artificial neural networks which can be trained to approximate the target probability distribution. In such situation the next sample can be generated independently by the neural network instead of relying on local updates of the previous sample. If the probability distribution parametrized by the neural network is sufficiently close to the target probability distribution most of the proposals would be accepted and the autocorrelations would be reduced. This algorithm, called Neural Markov Chain Monte Carlo (NMCMC), was studied in details in Ref.~\cite{Bialas:2021bei} using the two-dimensional Ising model as the testbed. It was confirmed that the critical slowing down can be much milder than in the local Metropolis algorithm \cite{metropolis}. Similar conclusions concerning the NMCMC autocorrelation times, in systems with continuous degrees of freedom, were reported earlier in Ref. \cite{Kanwar:2020xzo}.

The work of Ref.~\cite{2020PhRvE.101b3304N} was based on a previous observation that autoregressive neural networks can be applied to estimate extensive observables in statistical physics, in particular the free energy, which can be directly related to the Kulback–Leibler divergence, see \cite{2019PhRvL.122h0602W} for details. More recently that observation, reformulated in terms of normalising flows in order to accommodate continuous degrees of freedom, was applied to the popular elementary particle model of scalar $\phi^4$ field theory in two-dimensions in Ref.~\cite{PhysRevLett.126.032001}. This may be considered as true progress since estimating the free energy of a system with discrete or continuous degrees of freedom is a notoriously difficult problem for standard Monte Carlo approaches. Another application of autoregressive neural networks was reported in Ref.~\cite{Luo:2021ccm}.

Staying in the context of systems with discrete degrees of freedom, the main obstacle in scaling the NMCMC approach or the variational estimation of the free energy, to system sizes relevant from the point of view of particle or statistical physics is the numerical cost of the approach which dramatically increases with the system size. Most of the published results used the maximal lattice of $L \times L = 16 \times 16$ degrees of freedom. In this work we describe a mapping of the physical degrees of freedom to the inputs of a set of neural networks which considerably reduces the numerical cost. 
We demonstrate our approach on the two-dimensional Ising model where we simulate a system of $128 \times 128$ spins as a test of validity of our idea.
Compared to the original approach described in Ref.~\cite{2019PhRvL.122h0602W} which is valid for any system with discrete degrees of freedom, our algorithm is applicable only to systems with nearest-neighbour interactions and it relies on the well-known Hammersley-Clifford theorem \cite{Hammersley-Clifford, Clifford90markovrandom} and Markov field property of the Ising model. 

As suggested by Ref.~\cite{2019PhRvL.122h0602W}, the prohibitive scaling might be eliminated by replacing fully-connected layers by convolutional ones in the employed architecture of neural network. Convolutional layers proved their superiority in applications where the translational symmetry is expected on the level of input data. This is, however, not the case in the context of autoregressive networks discussed here, as neurons represent conditional probabilities of consecutive physical degrees of freedom and there is no symmetry on the level of conditional probabilities. In particular, the association of physical spins to consecutive neurons is arbitrary and can even be random. This implies that the neighbouring neurons do not necessarily correspond to spins which are close to each other on the physical lattice. This means that convolutional layers do not enforce the translational symmetry any more in this case. In fact they may even break it. Fully connected layers also do not enforce translational symmetry, but at least in theory can be trained to respect it. Note that in convolutional architectures the number of layers must scale like $L$ to fully account for all dependencies in conditional probabilities. It was shown in Ref.~\cite{2019PhRvL.122h0602W} that fully connected architectures are much faster than convolutional layers when actual run times are compared. Hence, in the following we do not consider convolutional layers as a viable solution for this particular class of problems.

As announced, we present our idea using the two-dimensional ferromagnetic Ising model with $N=L^2$ spins which are located on a $L\times L$ periodic lattice. The system is fully characterized by the Hamiltonian operator which provides the energy of a given configuration of spins $\mathbf{s} = \{ s^i \}_{i=1}^N$
\begin{equation}
    H(\mathbf{s}) = - J \sum_{\langle i,j \rangle} s^i \, s^j \,,
\label{Ising_hamilt}
\end{equation}
where the sum runs over all neighbouring pairs of lattice sites. We set the coupling constant $J=1$. Each configuration $\mathbf{s}$ appears with the Boltzmann probability
\begin{align}
    p(\mathbf{s}) = \frac{1}{Z} \exp(-\beta H(\mathbf{s})) \,,
    \label{eq_boltz_distr}
\end{align}
where $Z$ is the partition function $Z=\sum_{\mathbf{s}} \exp(-\beta H(\mathbf{s}))$ and the sum is performed over all $2^N$ configurations.

\section{Conditional probabilities and autoregressive networks}

In the approach discussed in earlier works \cite{2019PhRvL.122h0602W, 2020PhRvE.101b3304N} a neural network was proposed to model the probability $p(\b{s})$. To this end $p(\mathbf{s})$ was rewritten as a product of conditional probabilities of consecutive spins,
\begin{equation}\label{eq:conditional_probabilities}
    p(\mathbf{s}) = p(s^1)\prod_{i=2}^N p(s^i| s^1, \dots, s^{i-1})
\end{equation}
The neural network is used to model the conditional probabilities $p(s^i|\cdot)$ by approximating it by $q_{\theta}(\mathbf{s})$. Then,
\begin{equation}\label{eq:conditional_probabilities}
    q_{\theta}(\mathbf{s}) =  q_{\theta}(s^1)\prod_{i=2}^N q_{\theta}(s^i| s^1, \dots, s^{i-1}),
\end{equation}
where $q_{\theta}$ are quantities returned by the neural network.
In the simplest realization, the neural network has $N$ input and output neurons and the value of the $i$-th output neuron is interpreted as $q_{\theta}(s^i = +1 | s^1, s^2 , \dots, s^{i-1} )$, where we explicitly marked the dependence of $q$ on the internal parameters $\theta$ of the neural network. 
In order to ensure that the conditional probabilities depend only on the values of the preceding spins one uses autoregressive neural network, where the neural connections are only allowed to the neurons associated to the preceding spins. 

Knowing the target probability of the spin configuration, $p(\mathbf{s})$ and its value approximated by the neural network, $q_{\theta}(\mathbf{s})$, the training of the neural network is performed by minimizing the inverse Kullback–Leibler (KL) divergence,
\begin{equation}\label{eq-KL}
    D_{KL}(q_{\theta}|p) = \sum_{\mathbf{s}} \, q_{\theta}(\mathbf{s}) \left(\log q_{\theta}(\mathbf{s})-\log p(\mathbf{s})\right)= E[\log q_{\theta}(\mathbf{s})-\log
    p(\mathbf{s})]_{q_{\theta}(\mathbf{s})}.
\end{equation}
As we usually do not know the normalization factor $Z$, by inserting the target probability $p$ \emph{without} the normalization factor $1/Z$ into
the inverse KL divergence $D_{KL}(q_{\theta}|p)$ we get
\begin{equation}\label{eq:shifted-DKL}
\begin{split}
   F_{q} = E[\log q_{\theta}(\mathbf{s})-\log p(\mathbf{s}) -\log Z)]_{q_{\theta}(\mathbf{s})}= F + D_{KL}(q_{\theta}|p),
\end{split}
\end{equation}
where $F=-\log Z$ is the true free energy which does not depend on the neural network weights $\theta$, and $F_q$ its variational approximation. Hence,  minimization of $D_{KL}(q_{\theta}|p)$ is equivalent to pushing the value of $F_q$ towards the true value $F$ from above. This approach of using neural networks as a  functional ansatz for variational minimization is known as variational autoregressive networks (VAN) \cite{2019PhRvL.122h0602W}. During the training $F_q$ can approximated by the arithmetic average over $n$ (called \emph{batch size}, for which we use the value 1024) samples generated from the distribution $q_{\theta}$,
\begin{equation}\label{eq:Fq-autoregressive}
    \begin{split}
    F_q\approx \hat F_q =\frac{1}{n} \sum_{i=1}^n \left(\log q_{\theta}(\mathbf{s}_i)+\beta H(\mathbf{s}_i) \right),\quad \mathbf{s}_i \sim q_{\theta}.
\end{split}
\end{equation}
This quantity is minimized during the network training.
Its gradient estimator is calculated as \cite{2019PhRvL.122h0602W}
\begin{equation}
    \diff{\hat F_q}{\theta} 
    =\frac{1}{n} \sum_{i=1}^n \diff{ \log q_\theta(\mathbf{s}_i)}{\theta}  \left[ (\log q_\theta(\mathbf{s}_i) +\beta H(\mathbf{s}_i))-
    \frac{1}{n} \sum_{j=1}^n(\log q_\theta(\mathbf{s}_j) +\beta H(\mathbf{s}_j))
    \right]
    , \quad \mathbf{s}_i \sim q_{\theta}.
    \label{DKL_grad}
\end{equation}
Network weights are updated using Adam optimizer\cite{Adam}.

The generation of samples is done via {\em ancestral sampling}. We generate spins one by one starting from $s^1$. Once spins $s^1,\ldots,s^{i-1}$ are generated they are used as an input to the network to obtain the conditional probability $q_\theta(s^i=+1|s^1,\ldots,s^{i-1})$ according to which the spin $s^i$ is generated and the process is continued. Clearly this requires calling the network $L^2$ times and is the biggest drawback of this approach. As we are unable to reduce this number, in this contribution we limit ourselves to the reduction of the workload required by a single call of the network. 

In order to understand the scaling of the numerical cost of the VAN approach and compare it to our approach we take as our measure two numbers: the approximate number of trainable parameters of the autoregressive neural network which is proportional to the memory utilization and the approximate number of floating point operations needed to generate one configuration of spins which characterizes the numerical effort.  The actual run times, which depend on the implementation and on the hardware used, are discussed in Section \ref{sec: results}. In order to simplify the discussion we assume that neural networks are composed of a single fully-connected layer as this is sufficient for our aims of comparing the two approaches, i.e. including $m_l$ layers approximately multiplies the number of parameters and of floating point operations by $m_l$. We further assume that this layer has same number of input and output neurons i.e. it looks like the lower layer in Fig.~\ref{fig:net}. Adding another layer with smaller number of output neurons does not change our conclusions about scaling. In what follows we consider large $L\gg 1$ and we keep only the leading power of $L$ which is denoted with the "$\approx$" sign.

The original proposal of Ref.~\cite{2019PhRvL.122h0602W} assumed a single 
fully-connected neural network for all spins and a mask in order to impose autoregressivity, which 
corresponds to $L^4+L^2+L^2 \approx L^4$ trainable parameters. The first term counts the weights, the second corresponds to the bias on each neuron and the third counts the parameters in the \verb[PReLU[ activation function used in that Reference. In principle, one could exploit the triangular (autoregressive) structure of the weight matrix to reduce the number of parameters by $\frac{1}{2}$. Additionally, one could as well implement another optimization such as the chessboard factorization \cite{Bialas:2021bei} and further reduce the scaling by $L^2 \rightarrow \frac{1}{2} L^2$, giving in total $\frac{1}{8}L^4+L^2$. Nevertheless, the number of trainable parameters would still increase with a leading scaling power of $L^4$. As far as the number of floating point operations is concerned, in order to generate one configuration of spins we need to invoke the neural network $L^2$ times (number of spins on the lattice) and each invocation has of the order of $L^4$ floating point operations. Hence, in total the number of operations needed to generate of one configuration scales as $L^6$ in the VAN approach.

In the following we propose to reorganize the neural network in such a way as to keep the same expressivity, but reduce the number of parameters and hence considerably reduce the numerical cost of the approach allowing us to simulate systems of size 64 times larger than the original proposal. With much smaller neural networks our idea considerably reduces memory requirements.

\section{Boundary and interior spins}

\begin{figure}
    \begin{center}
    \includegraphics[width=0.3\textwidth]{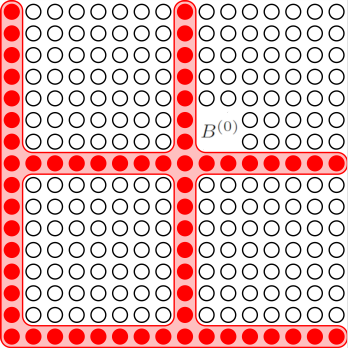}\hspace{3mm}%
    \includegraphics[width=0.3\textwidth]{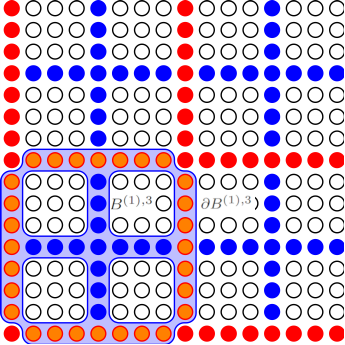}\hspace{3mm}%
    \includegraphics[width=0.3\textwidth]{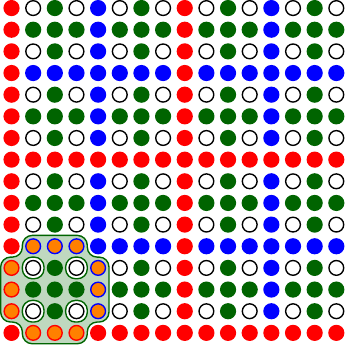}
    \end{center}
    \caption{Example of hierarchical partitioning for $L=16$. On the first level, the red boundary $B^{(0)}$ highlighted on the left panel of the figure is generated with one neural network $\mathcal{N}^0$. At the second level of hierarchy, one neural network of smaller size $\mathcal{N}_1$ is used to consecutively fix four sets of boundaries shown in blue. The example of $B^{(1),3}$ is highlighted in the middle panel. The surrounding spins $\partial B^{(1),3}$ are shown in orange. At the third level of hierarchy, one neural network of even smaller size $\mathcal{N}_2$ is used to consecutively fix sixteen sets of boundary spins marked in green. The example of $B^{(2),15}$ is highlighted in green on the right panel. The remaining empty spins corresponding to $I^{k} \equiv B^{(3),k}$, $k=1,\dots,64$ have all the neighbours fixed and therefore can be generated from a local Boltzmann distribution with the heatbath algorithm.}
    \label{fig:sublattices4}
\end{figure}

\begin{figure}
\begin{center}
   \includegraphics[width=0.9\textwidth]{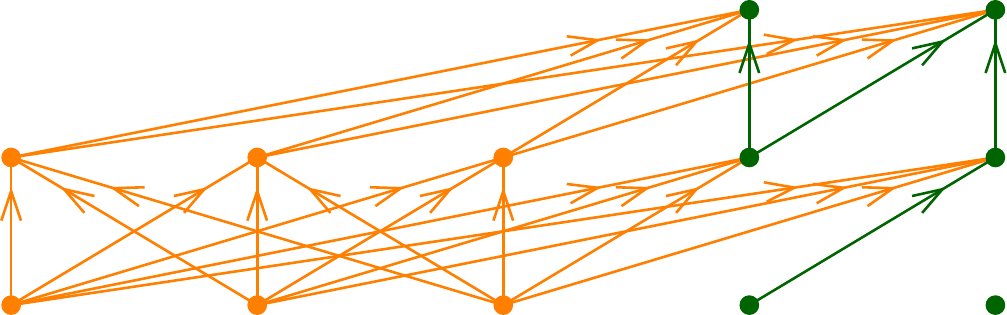} 
\end{center}
\caption{\label{fig:net}Scaled down representation of the architecture of the smallest neural network used to generate green sublattice (Figure~\ref{fig:sublattices4} right). Orange nodes correspond to border spins and green nodes to interior spins. Orange nodes are connected to every node of the subsequent layer. Green nodes are connected in an autoregressive way. The propagation of signals starts at the bottom of the figure and goes upward. The real network used in this case has nine orange input neurons and five green.} 
\label{fig:nn}
\end{figure}

We propose to divide the spins into two species: boundary spins and interior spins, as demonstrated in  Fig.~\ref{fig:sublattices4} (left) for periodic lattice. For non-periodic lattice the only difference is the shape of the red boundary spins. In the presented partitioning of a $16 \times 16$ lattice we obtain 60 red boundary spins and four sublattices with 49 white interior spins each. Each sublattice has boundaries fixed by a subset of red boundary spins. 
The main advantage of such a partitioning is the following fact which can be easily proven and is known as the Hammersley-Clifford theorem: for each sublattice, once the boundary made of red spins is fixed, the probabilities of interior spins {\em do not depend}  on the values of the spins outside the surrounding red boundary. This allows us to write $p(\mathbf{s})$ as the product
\begin{equation}
    \label{eq:conditional_probabilities3}
    p(\mathbf{s}) = p(\textrm{B}(\mathbf{s})) p(\textrm{I}(\mathbf{s})| \textrm{B}(\mathbf{s})) = p(\textrm{B}(\mathbf{s})) \prod_{a=1}^{4} p(\textrm{I}^a(\mathbf{s})| \textrm{B}^a(\mathbf{s})),
\end{equation}
where the operation B, $\textrm{B}(\mathbf{s})$, extracts the boundary spins from the configuration $\mathbf{s}$, whereas I, $\textrm{I}(\mathbf{s})$ the interior spins. We explicitly indicated that the probabilities of interior spins are conditional probabilities depending on the boundary spins. Each individual factor can be further factorized as
\begin{equation}\label{eq:conditional_probabilities2}
    p(\textrm{B}(\mathbf{s})) = p(s_B^1)\prod_{i=2}^{N_B} p(s_B^i | s_B^1, s_B^2, \dots, s_B^{i-1})
\end{equation}
and
\begin{equation}\label{eq:conditional_probabilities2p}
    p(\textrm{I}^a(\mathbf{s}) | \textrm{B}^a(\mathbf{s})) = \prod_{i=1}^{N_I} p(s_I^{a,i} | s_I^{a,1}, s_I^{a,2}, \dots, s_I^{a,i-1};  \textrm{B}^a(\mathbf{s}) ).
\end{equation}

At this point we can use \emph{two} neural networks: one for the boundary spins and one for the interior spins. The neural network for the boundary spins is of the same type as in the original NMCMC algorithm, but considerably smaller, its width is $4L-4$ and hence the number of its trainable parameters scales as $L^2$. The second type of neural network used to parametrize the probabilities of the interior spins  has $(\frac{1}{2}L-1)^2$ neurons corresponding to interior spins (white) and $2L-4$ neurons corresponding to boundary spins (red), resulting in total of $\approx \frac{1}{4}L^2$ of input neurons.
This network has a fully connected part from the boundary input neurons (orange) to all output neurons, and an autoregressive part from interior input neurons to interior output neurons (green) (see  Figure~\ref{fig:net}). 
Note that the {\em same} neural network is needed for all 4 sublattices. This may be interpreted as introducing a substitute of translational symmetry into the approach. However, we implement it on the level of physical degrees of freedom where this symmetry is defined, and not on the level of conditional probabilities, as would convolutional layers do. 
Calculating the total number of trainable weights in this setup we still get the unfavorable scaling of $L^4$ which originates from the neural network used to generate the interior spins in the sublattices. However, one can easily improve on this, by applying the division into boundary and interior spins to each sublattice. This will be described in the next Section.

\section{Hierarchical Autoregressive Networks (HAN)}

Based on the conclusion from the preceding Section, we describe an  algorithm for spin generation based on iterative division of the lattice into smaller sublattices.  Below we describe the procedure and provide a pseudocode which illustrates the algorithm.

We start again with the division of spins into boundary and interior spins (see the left panel of Fig.~\ref{fig:sublattices4}). The first set of boundary spins is marked in red and we generate them as described in the previous Section. We denote these spins by $B^{(0)}$ where the upper index corresponds to the zeroth step of consecutive iterations consisting of division and fixing of spins. Similarly, the corresponding autoregressive neural network trained to fix these spins is denoted by $\mathcal{N}^0$. The spins in $B^{(0)}$ do not depend conditionally on any other variables. Now, once those spins are fixed, we take each sublattice and treat it as a new problem with fixed boundaries. Instead of directly generating all interior spins, we iterate and divide it further into boundary and remaining spins. The sets of boundary spins $B^{(1),k}$ on each of the sublattices are marked in blue in the middle panel of Fig.~\ref{fig:sublattices4}. One example of a new set of spins at the second level of iteration, $B^{(1),3}$, is highlighted . We propose to use another autoregressive neural network, called $\mathcal{N}^1(\partial B^{(1),k})$ to generate the blue boundary spins. The main advantage of our approach is that one network $\mathcal{N}^1$ is used to generate all $B^{(1),k}, \, k=1,2,3,4$. The probabilities of blue spins depend conditionally on the surrounding spins, we denote the latter by $\partial B^{(1)}$. The middle panel in Fig.~\ref{fig:sublattices4} shows the example of $B^{(1),3}$ and highlights in orange  the surrounding spins belonging to $\partial B^{(1),k}$. The geometry of the neural network which must be used in this case has a fully connected part, an autoregressive part, with the values of the surrounding spins playing the role of external parameters which is explicitly marked by the argument of the neural network $\mathcal{N}^1$. We show an example of such neural network in Fig.~\ref{fig:net}. Repeating this procedure iteratively, we stop when $\frac{3}{4}L^2$ boundary spins (at different levels of iterations) are generated and the remaining $\frac{1}{4}L^2$ interior spins can be set by the heatbath step\footnote{In the heatbath algorithm we draw a spin from probability distribution fixed by the values of its neighbours: probability of spin $s^i$ to be $+1$ or $-1$ is equal $q(s^i=\pm 1, n(s^i))=  \left[ 1+\exp \left(\mp\, 2\beta  \sum_{\ j \in n(s^i) } s^j \right) \right]^{-1}$, where the sum is performed over neighbours $n(s^i)$.} (see the the right panel of Fig.~\ref{fig:sublattices4}). The heatbath algorithm may in principle be replaced by a further level in the hierarchy of neural networks. However, since that single spin can be generated with exact probability, we implement that optimization directly. We summarize these steps with the pseudocode presented in Algorithm \ref{alg. 1}.

\begin{algorithm}
\begin{center}
\caption{\label{alg. 1}Iterative generation of all spins in a single configuration.} 
\begin{algorithmic}[0]
 \Procedure{HAN}{$\beta$} \Comment{generate spin configuration $C$ at given $\beta$}
      \State define $B^{(0)}$
      \State $B^{(0)} \gets \mathcal{N}^0(\textrm{void})$
      \Comment{spins $B^{(0)}$ are set by $\mathcal{N}^0$ without any additional dependence}
      \State $C \gets B^{(0)}$
      \For{$m=1,\dots,\log_2 L$}
        \For{$k=1,\dots,4^m$}
            \State define $B^{(m),k}$ \Comment{one set of blue sites on Fig.\ref{fig:sublattices4} (middle)}
            \State define $\partial B^{(m),k}$ \Comment{set of surrounding spins of $B^{(m),k}$}
            \State $B^{(m),k} \gets \mathcal{N}^m(\partial B^{(m),k})$ \Comment{spins $B^{(m),k}$ are set by $\mathcal{N}^m$ depending on $\partial B^{(m),k}$}
            \State $C \gets B^{(m),k}$
        \EndFor
      \EndFor
      \Comment{only single spins remain unset, see white sites on Fig.\ref{fig:sublattices4} (right)}
      \State $I \gets $ HB$(\beta, C)$ \Comment{fix remaining spins with the heatbath algorithm}
      \State $C \gets I$ 
      \Comment{all spins are fixed}
      \State \textbf{return} $C$
\EndProcedure
\end{algorithmic}
\end{center}
\end{algorithm}



\
In order to bound the numerical cost of our approach, we again count the total number of trainable parameters needed to describe a system of extent $L=2^m = l^{(0)}$, $m>1$. Again, we only keep the leading powers of $L$, omitting the parameters hidden in the bias and activation functions. The first step differs from the recursively defined division into boundary and interior spins as it needs to account for the periodic boundary conditions of the physical system. We already discussed that step and obtained $N_p^{(0)} = (4L-4)^2 \approx 16 L^2 = 2^{2m+4}$ parameters. The resulting sublattices have linear extent of $l^{(1)} = \frac{1}{2}(l^{(0)}-2) = 2^{(m-1)}-1$. 

If $m>2$, we can  continue this process and divide each sublattice by  generating first a new set of boundary spins  (blue spins in Fig.~\ref{fig:sublattices4} middle) that divide each sublattice in four sublattices. To this end we use  a new neural network.  All blue spins in given sublattice depend conditionally on $4l^{(1)}$ red boundary spins which surrounds them (orange spins in Fig.~\ref{fig:sublattices4} middle), as already explained previously. We have $2l^{(1)} -1$ blue spins and hence the fully connected dense layer will have $N_p^{(1)} \approx (6l^{(1)} -1)^2 \approx 9\ 2^{2m}$ parameters. At each subsequent recursive partitioning, we find a sublattice of linear size
\begin{equation}
    l^{(i)} = \frac{1}{2} \big( l^{(i-1)} -1 \big) = 2^{m-i} - 1
\end{equation}
and the number of trainable parameters of network $\mathcal{N}^i$ is
\begin{equation}
    N_p^{(i)} \approx (6 l^{(i)}-1)^2 \approx 9 \ 2^{2(m-i+1)}.
\end{equation}
The recursion ends when $l^{(i)} = 3$, hence $i=m-2$. Gathering the individual contributions from each iteration we finally obtain
\begin{equation}
    N_p = \sum_{i=0}^{m-2} N_p^{(i)} \approx 
    2^{2m+4} + 9\sum_{i=1}^{m-2} 2^{2(m-i+1)}\approx 28 L^2
    \label{eq. result}
\end{equation}
Therefore, the scaling of parameters number with $L^4$ present in the original approach was completely eliminated and replaced with $L^2$. 

The number of floating point operations for the neural network at $i$th level of the hierarchy is a product of $N_p^{(i)}$ and number of network invocations (note that for each spin we need one invocation of network):
\begin{align}
    C^{(0)} &= (2^{m+2}-4) N_p^{(0)} \approx  64 \ 2^{3m}, \\
    C^{(i)} &= 4^i (2 l^{(i)}-1) N_p^{(i)} \approx  72 \  2^{3m-i}.
\end{align}
Summing up these numbers we obtain the number of floating point operations for one layer,
\begin{equation}
    C = C^{(0)} + \sum_{i=1}^{m-2} C^{(i)} \approx 136 L^3,
    \label{eq. result numerical}
\end{equation}
compared to the $\propto L^6$ operations in the VAN approach.

 The numerical cost of calculating the probability $q(\mathbf{s})$ of a configuration which has been already generated is given by,
\begin{equation}
    C_q = N_p^{(0)} + \sum_{i=1}^{m-2} 4^i N_p^{(i)} \approx 36L^2\log_2 L,
    \label{num_cost_q}
\end{equation}
since neural networks are invoked only once for each $B^{(i),k}$.

\section{Numerical results}
\label{sec: results}

\begin{figure}
     \centering
     \includegraphics[width=0.65\textwidth]{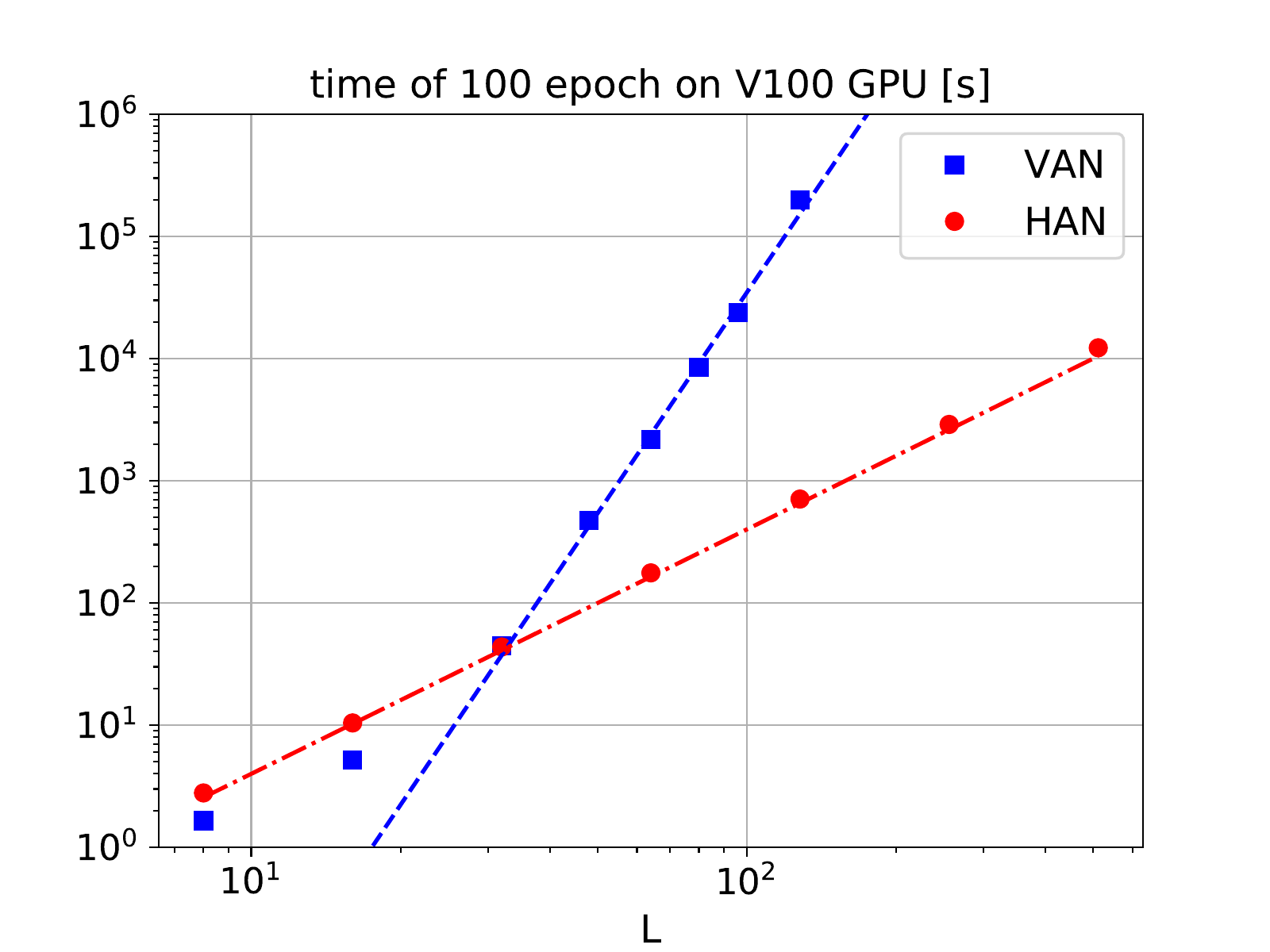}
     \caption{Time scaling of the VAN and HAN algorithms: we show time of training for 100 epochs in both approaches as function of system size $L$. Approximate scaling as $L^2$ for HAN and $L^6$ for VAN can be seen for sufficiently large values of $L$ (lines).}
     \label{fig:scaling}
 \end{figure}

In the numerical experiments we used the an adaptation of the \verb[PyTorch[ implementation of autoregressive layers shared by the authors of Ref.~\cite{2019PhRvL.122h0602W}. All neural networks $\mathcal{N}^i$ consist of two densely connected layers with some connections removed. The architecture was schematically shown in Fig.~\ref{fig:nn}. The neural network used to model the first boundary spins does not have orange nodes. First and second layers use \verb[PReLU[ and \verb[sigmoid[ activation functions respectively. 

As a demonstration of the practical impact of our proposal we compared our approach with the results obtained using standard VAN from Ref.~\cite{2019PhRvL.122h0602W}. For a given size of the spin lattice we measured the time needed to perform 100 epochs of training of a single neural network (VAN) and the hierarchy of autoregressive networks (HAN) in the new approach. We plot the results obtained using a single V100 GPU in Fig.~\ref{fig:scaling}. We checked that the time of all additional operations is negligible compared to the neural network evaluations.
We notice that for small values of $L$, $L\le16$, VAN approach is faster and the time of both methods is comparable for $L=32$. For larger systems the VAN approach scale approximately as $L^6$. As for the HAN approach we clearly see a $L^2$ scaling. We checked that these times are saturated by the multiple invocations of the smallest neural networks. Hence, the visible $L^2$ scaling can be associated with the overhead of numerous GPU kernel invocations (which indeed increases as $L^2$) and hides the expected numerical effort scaling. Also for very small neural networks where the number of parameters is smaller then the number of ALUs on the GPU (5120 for the V100) we expect that time for single network invocation independent on network size. Decreasing the number of GPU calls should further reduce the run time of the HAN approach. Still, for $L>32$ already at this stage our proposal yields much faster configuration generation and training.

What we also expect is that the quality of training is much better than in the variational approach from Ref.~\cite{2019PhRvL.122h0602W}. Our recursive model for approximating the Boltzmann probability distribution of spin configurations should allow to obtain a variational estimate of the free energy which is much closer to the exact value $F$. The latter is obtained using known analytical expressions for Ising model for given $\beta$ and $L$, see Refs.~\cite{PhysRev.185.832, 2020PhRvE.101b3304N}.

\begin{figure}
     \centering
     \includegraphics[width=0.48\textwidth]{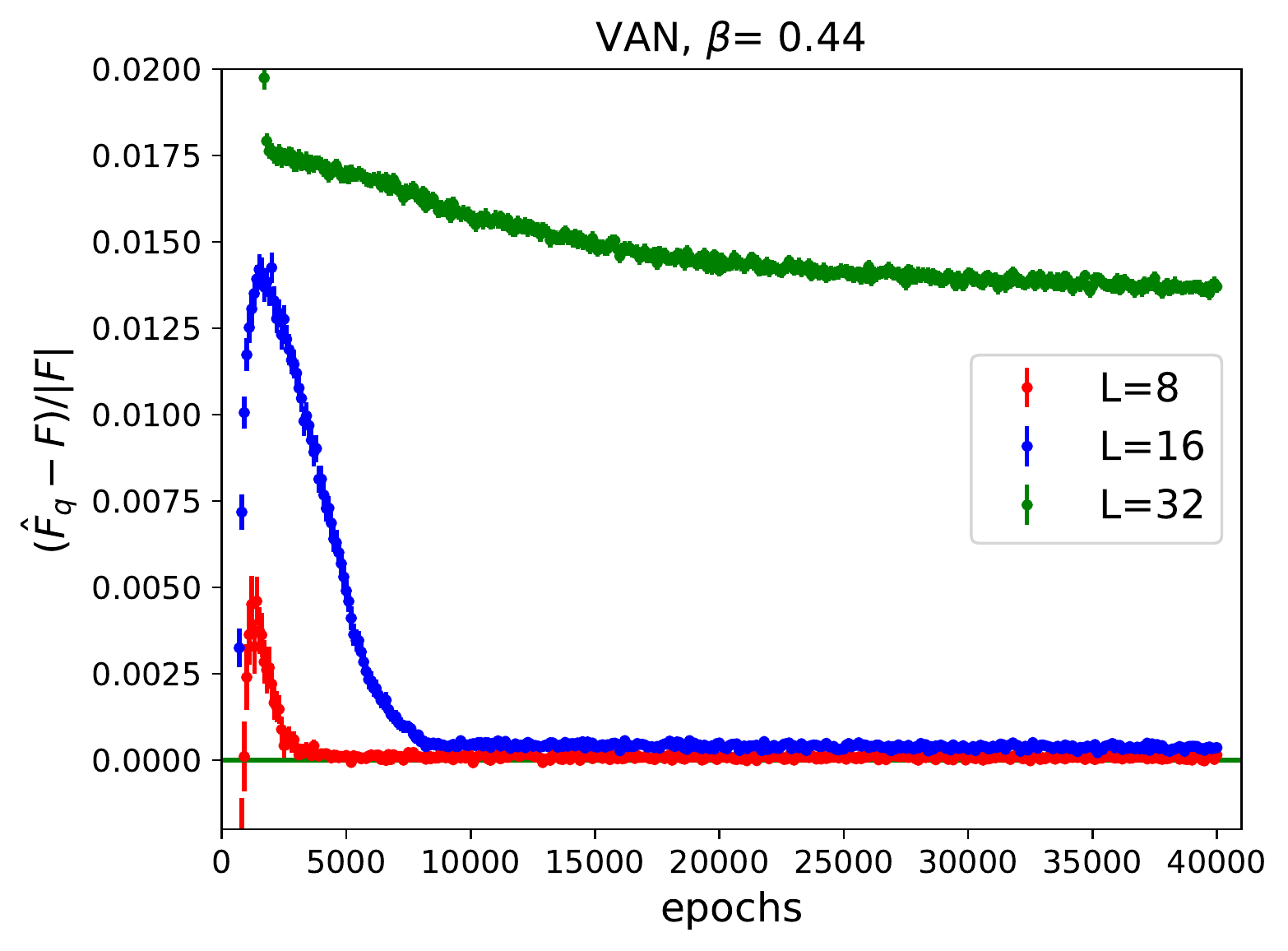}
     \includegraphics[width=0.48\textwidth]{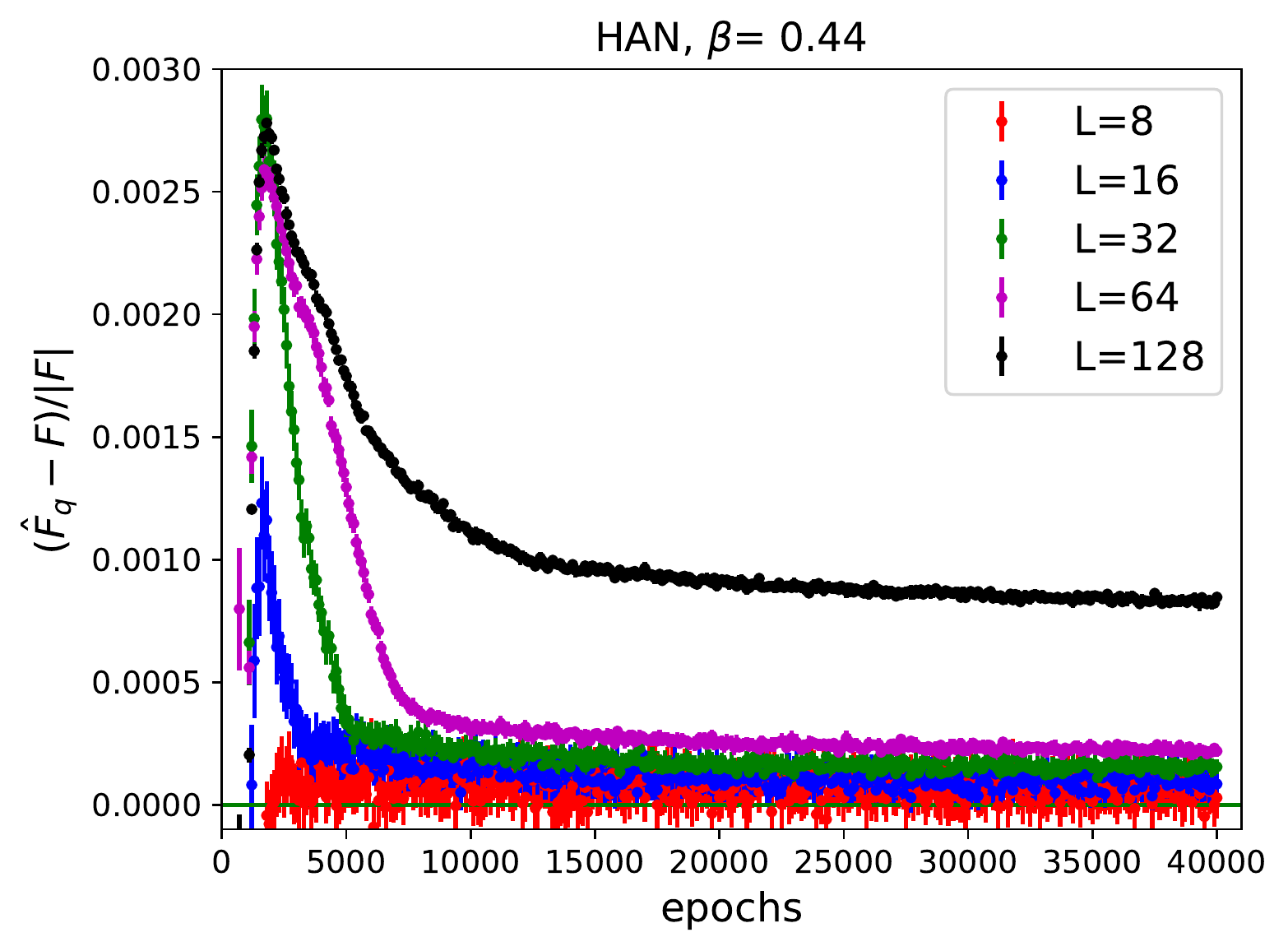}
     \caption{Relative deviation of variational free energy $\hat F_q$ from exact value as a function of the neural network training time on lattices from $L=8$ up to $L=128$ obtained with the original and hierarchical algorithm. A large improvement can be observed for the new algorithm, as the precision of $\hat F_q$ is below 0.1\%.}
     \label{fig:results1}
 \end{figure}

In order to illustrate this we show in Fig.~\ref{fig:results1} the relative difference between the estimated free energy and exact value, $(\hat F_q-F)/|F|$. We plot it for several values of $L$ and $\beta=0.44$, where the expected phase transition is approximately located. For both approaches the same $\beta$-annealing algorithm was applied: during training the beta was changed according to $\beta(epoch)=0.44(1-0.996^{epoch})$. We checked that the use or not of $\beta$-annealing does not effect the final results for the free energy. $\hat{F}_q$ is estimated from the batch during training and is plotted together with its statistical uncertainty. In the original VAN (left panel in Fig.~\ref{fig:results1}) the training of the neural network needed to model the system of size $L=32$ had difficulties to converge and yielded an approximation of the free energy 1.5\% away of the expected result. On the contrary, the HAN approach offers faster training and the discrepancy between the approximated free energy and the expected value below 0.1\% even for the system of size $L=128$ (see right panel in Fig.~\ref{fig:results1}).

The results for $\hat{F}_q$ shown in Fig.~\ref{fig:results1} have been obtained with the $Z_2$ symmetry (present in the Ising model) imposed during training as explained in Ref.~\cite{2019PhRvL.122h0602W}: for a sampled configuration $\mathbf{s}$, it requires replacement $q_\theta(\mathbf{s})\rightarrow \left[ q_\theta(\mathbf{s}) + q_\theta(- \mathbf{s}) \right]/2$, where $- \mathbf{s}$ is obtained from $\mathbf{s}$ by flipping all the spins. As it was mentioned in Ref.~\cite{2019PhRvL.122h0602W} inclusion of $Z_2$ significantly improves training efficiency. The effect of this symmetry was then further investigated in Ref.~\cite{Bialas:2021bei} by the present Authors, where we have shown that imposing the $Z_2$ leads to improvement of training only in the ordered state, namely for $\beta>0.44$. For the disordered state the imposition of another type of symmetries for the Ising model (with periodic boundary conditions) turns out to be beneficial -- translational symmetries:
\begin{equation}
    \mathcal{S}_i^{T_x} \,\mathbf{s}_{x,y} = \mathbf{s}_{x-i,y}, \quad \mathcal{S}_i^{T_y} \, \mathbf{s}_{x,y} = \mathbf{s}_{x,y-i}, \qquad 0 \le i < L,
\end{equation}
where $\mathbf{s}_{x,y}$ is a spin in the position $(x,y)$ on the $L\times L$ lattice. In Ref.~\cite{Bialas:2021bei} we have shown that imposing $T_y$ symmetry implemented by the replacement,
\begin{equation}
    q_\theta(\mathbf{s})\rightarrow \left[ q_\theta(\mathbf{s}) + q(\mathcal{S}_1^{T_y} \mathbf{s}) + \dots + q(\mathcal{S}_{L-1}^{T_y} \mathbf{s}) \right]/L,
    \label{Ty_symmetry}
\end{equation}
further improves training efficiency.
\begin{figure}
     \centering
     \includegraphics[width=0.55\textwidth]{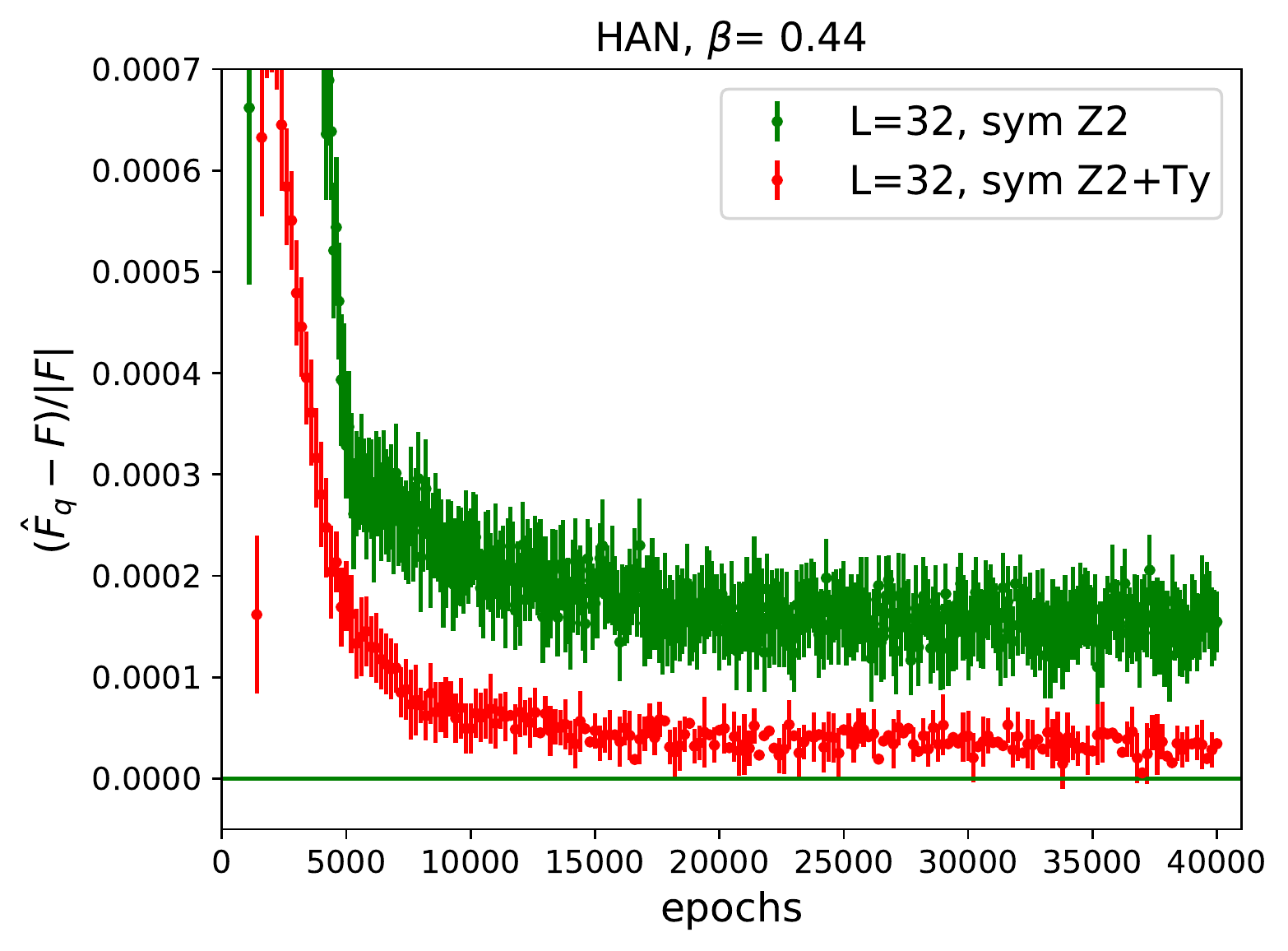}
     \caption{Effect of symmetries applied during the training:  variational free energy $\hat F_q$ as a function of the neural network training time on lattices size $L=32$ obtained when only $Z_2$ symmetry is applied and when one combines $Z_2$ and $T_y$ symmetries.}
     \label{fig:results_Z2+Ty}
 \end{figure}
We show the effect of this improvement in Fig.~\ref{fig:results_Z2+Ty} where the $T_y$ symmetry was included during the training. With this enhancement we reached the precision $5\times10^{-5}$ (red dots) for the system with $L=32$ compared with $0.002$ for HAN with $Z_2$ (green dots) and $0.015$ given by VAN with $Z_2$ (green dots in left panel of Fig.~\ref{fig:results1}).  The numerical cost of symmetry imposition adds up to the cost of configurations generation. Each additional calculation of $q_\theta(\mathbf{s})$ which cost is given by Eq.\eqref{num_cost_q} is subleading compared to the cost of configurations generation (\ref{eq. result numerical}) as long as the number of such additional invocations does not scale as $L$.
Translational symmetries are an exception here, because the number of possible transformations is given by $L$ in each of the dimensions. Hence, the imposition of the $T_y$ symmetry as indicated by Eq.~\eqref{Ty_symmetry} introduces an additional power of $L$ to the cost of the symmetry imposition which would grow like $\sim L^3\log_2 L$. Adding also $T_x$ symmetry would further change scaling into $\sim L^4 \log_2 L$.

As was already mentioned, both VAN and HAN approaches can be applied to not only to the variational approximation of the free energy but also as a generator of Markov Chain Monte Carlo propositions. Using neural networks saved after a training consisting of 40000 epochs we generated a Markov chain of length $N_{MC}$. As was shown in Fig.~\ref{fig:results1} this is the number of epochs after which we observe no further improvement in the training for both algorithms and all system sizes. The consecutive propositions in the Markov chain 
\begin{equation}
    \mathbf{s}_1, \ldots, \mathbf{s}_{N_{MC}},
\end{equation}
were accepted with probability
\begin{equation}
    \min \left( 1, \frac{p(\mathbf{s}_{k+1}) q_\theta(\mathbf{s}_{k})}{p(\mathbf{s}_{k}) q_\theta(\mathbf{s}_{k+1}) }   \right). 
\label{accept_rej_condition}
\end{equation}
This procedure guarantees that spin configurations are distributed according to $p$ distribution when $N_{MC}\rightarrow \infty$.
When $q_\theta\neq p$ some propositions are rejected, leading to repetition of configurations in MCMC and nonzero autocorrelation. In order to quantify them we define the normalized autocorrelation function for any observable $\mathcal{O}$ (with mean $\bar{\mathcal{O}}$ and variance $\operatorname{var} \mathcal{O}$) as
\begin{equation}
    {\Gamma}_\mathcal{O}(t) = 
    \frac{1}{\operatorname{var} \mathcal{O} }  \frac{1}{N_{MC}}\sum_{i=1}^{N_{MC}}\left( \mathcal{O}(\mathbf{s}_{i})-\bar{\mathcal{O}}\right) \left(\mathcal{O}(\mathbf{s}_{i+t}) - \bar{\mathcal{O}}\right),
\label{autocorr_func_practice}
\end{equation}
and the integrated autocorrelation time
\begin{equation}
     \tau^{int}_\mathcal{O} =1+2\sum_{i=1}^{t_{max}}  \Gamma_\mathcal{O}(t).
\label{iat_def_est}
\end{equation}
The sum is performed up to the value $t_{max}$, where ${\Gamma}_\mathcal{O}(t)$ becomes negative (due to statistical fluctuations). In what follows we choose energy $H$ given by Eq.~\eqref{Ising_hamilt} as the operator $\mathcal{O}$ and generate Markov chain of length $N_{MC}=6\cdot 10^5$ configurations. We use them to measure $\tau^{int}$ (note that we skipped $H$ subscript). We choose $\beta=0.44$, where $\tau^{int}$ is maximal due to the critical slowing down (see Ref.~\cite{Bialas:2021bei} for discussion of this phenomenon in the normal VAN approach). 

\begin{figure}
     \centering
     \includegraphics[width=0.65\textwidth]{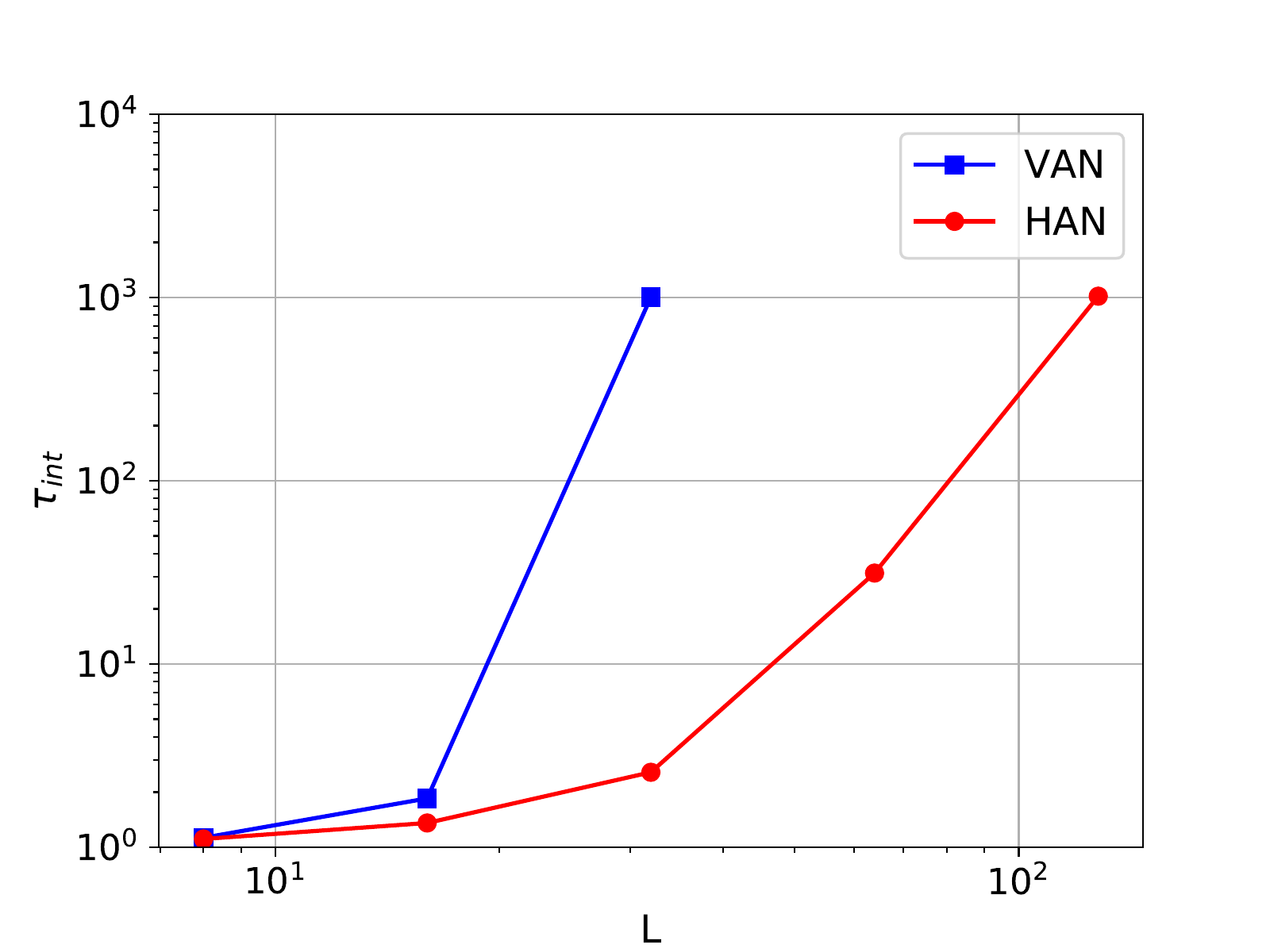}
     \caption{Integrated autocorrelation time of the hierarchical algorithm as a function of the system size $L$.}
     \label{fig:tau}
 \end{figure}

Fig.~\ref{fig:tau} compares the integrated autocorrelation time for VAN and HAN as a function of system size $L$. In both approaches we observe a rise of $\tau^{int}$ with $L$. However, it is much faster for the VAN approach. It can be seen that HAN for system $128\times 128$ achieves similar autocorrelation time as the VAN algorithm for $32\times 32$. These data did not include the effects of the imposition of the translational symmetry. Since this symmetry considerably improves the quality of the training of the neural network, we expect that the integrated autocorrelation times may be even smaller. However, due to the increased numerical cost of that symmetry we did not consider it here.


\section{Conclusions}

In this work we have described  a particular scheme of generating spin configurations using autoregressive neural networks which has a better scaling property than the reference approach. We have provided estimates of its numerical cost which showed that it scales with the system size as $L^3$ as opposed to $L^6$ of the VAN approach. The main improvement comes from the fact that we have replaced a single autoregressive neural network of width $\sim L^2$ by a hierarchy of $\log_2 L-1$ neural networks of decreasing width. It follows that the number of neural networks weights is much smaller in the new scheme, thus leading to a much smaller memory consumption. We expect that, from the practical perspective, the improvement is even larger because it usually takes less steps of the training algorithm to provide optimal results for a neural network with smaller number of weights. Our idea can improve both, the variational method of estimating the free energy of statistical systems, and the autocorrelation times in the  Markov Chain Monte Carlo simulations. Eventually, we emphasize that the hierarchical structure of neural networks allowed us to reintroduce translational invariance on the level of physical degrees of freedom. Further improvements may benefit from this fact.

\section{Outlook}

The following observations may serve as possible future research directions. 

First, in this work we have demonstrated that the improved scaling of our proposal works well for the two-dimensional Ising model. However, it is possible to generalize our recursive approach to higher dimensions as well as to systems with more complex interactions. For example spin glasses models where the coupling constant $J$ depends on the bound can also be simulated. In a straightforward manner the random values of $J_{ij}$ can be included in the  structure of neural networks on subsequent levels in the hierarchy as additional external parameters. In this way the conditional dependence on $J_{ij}$ can be incorporated allowing the neural network to correctly account for the functional dependence on $J_{ij}$. One may expect that if the training of such hierarchy of neural networks is successful, it could be performed once and used for any set of $J_{ij}$ without the need of re-training, in contrast to the VAN approach \cite{2019PhRvL.122h0602W} or alternative method suggested in Ref.~\cite{2020PhRvE.101e3312M}. Moreover, since the neural networks provide global updates, such approach might be able to overcome the notorious problem of traditional methods of getting trapped in one of the local minima of the free energy. Recent results showing that ergodic sampling in a case of complicated energy landscape in the context of normalizing flows was reported in Refs.~\cite{Hackett:2021idh,Albergo:2022qfi}. Hence, we expect significant improvement of the quality of results with the HAN approach applied to spin glass type of systems.

Second, the approach proposed in this work offers an improved training strategy, as the training of neural networks for larger spin systems does not need to be started from random weights. Indeed, small neural networks at the bottom of the HAN hierarchy can be trained to high accuracy using much smaller spin systems, their states can be saved and reused for larger spin systems. This is due to the fact that the conditional probabilities depend only on the surrounding spins and are independent of the total size of the simulated system. So in principle when doubling the system linear size $L$, only the two largest neural networks $\mathcal{N}^0$ and $\mathcal{N}^1$  have to be constructed and trained "from scratch". The rest of the hierarchy $\mathcal{N}^i$ can be reused from the smaller system. Those networks could be used as they are, without any further training, but probably better results can be achieved with some additional training which we would expect to converge faster.

One may also expect that smaller batch sizes can be used for training. One may observe that the smaller neural networks at higher iterations of the division are trained using multiple sublattices of the original configuration. In fact at iteration $i$, we have $4^i$ sublattices which we invoke in the training process and hence the variance of the gradient is $2^i$ smaller than in the original approach. Hence, we expect that the quality of training of these inner neural networks will be very high. In the case when such precision is not necessary, one may try to reduce the batch size.

 Eventually, the natural question which arises, whether the hierarchical generation presented in this work can be generalized to systems with continuous degrees of freedom. Our preliminary studies show that this can be done for the scalar $\phi^4$ field theory using Normalizing Flows. However, using the hierarchy of neural networks with dense layers turns out to be not as beneficial as we have demonstrated in the case of autoregressive networks. Further studies are needed for more definite conclusions.

\section*{Acknowledgments}
Computer time allocation 'plgtmdlangevin2' and 'plgnnformontecarlo' on the Prometheus supercomputer hosted by AGH Cyfronet in Krak\'{o}w, Poland was used through the polish PLGRID consortium. T.S. kindly acknowledges support of the Polish National Science Center (NCN) Grants No.\,2019/32/C/ST2/00202 and 2021/43/D/ST2/03375 and support of the Faculty of Physics, Astronomy and Applied Computer Science, Jagiellonian University Grant No.\,2021-N17/MNS/000062. This research was partially funded by the Priority Research Area Digiworld under the program Excellence Initiative – Research University at the Jagiellonian University in Kraków.




\bibliography{references2}


\end{document}